\documentclass[conference]{IEEEtran}
\IEEEoverridecommandlockouts
\usepackage{cite}
\usepackage{amsmath,amssymb,amsfonts}
\usepackage{algorithmic}
\usepackage{graphicx}
\usepackage{textcomp}
\usepackage{xcolor}
\usepackage[hyphens]{url}

\usepackage{array}
\usepackage{booktabs}
\usepackage{multirow}
\usepackage{subfig}
\usepackage{tabularx}
\usepackage[dvipsnames,table]{xcolor}
\usepackage{fontawesome}
\usepackage{pifont}
\usepackage{tikz}

\newcommand{\cmark}{\textcolor{Green}{\ding{51}}}
\newcommand{\xmark}{\textcolor{Red}{\ding{55}}}
\newcommand{\halfmark}{\textcolor{YellowOrange}{\faCheckCircleO}}

\newcommand{\mypara}[1]{\vspace{2pt}\noindent{\textbf{#1}}}

\def\BibTeX{{\rm B\kern-.05em{\sc i\kern-.025em b}\kern-.08em
    T\kern-.1667em\lower.7ex\hbox{E}\kern-.125emX}}
\begin{document}

% Ensure letter paper
\pdfpagewidth=8.5in
\pdfpageheight=11in

%%%%%%%%%%%---SETME-----%%%%%%%%%%%%%
\newcommand{\iscasubmissionnumber}{NaN}
%%%%%%%%%%%%%%%%%%%%%%%%%%%%%%%%%%%%

\pagenumbering{arabic}

%%%%%%%%%%%---SETME-----%%%%%%%%%%%%%
\title{Reconfiguring System-Level Protection Through Bounded Delegation in RISC-V MCUs}

\author{Luís Cunha, José Martins, Manuel Rodríguez, Tiago Gomes, Uwe Moslehner, \\ Kai Dieffenbach, Georg Schmalz, Kajetan Nuernberger, Thomas Roecker, and Sandro Pinto
\thanks{This work has been supported by FCT – Fundação para a Ciência e Tecnologia within the R\&D Unit Project Scope UID/00319/2025 - Centro ALGORITMI (ALGORITMI/UM) https://doi.org/10.54499/UID/00319/2025 and the Scholarships Project Scope SFRH/BD/01007/2023. This work is funded in part by the RIGOLETTO project supported by the EU Horizon Europe research and innovation program under Grant Agreement Number 101194371; and partially supported by the Federal Ministry of Research, Technology and Space under the funding code 16MEE0548S.}%
\thanks{Luís Cunha, Manuel Rodríguez, Tiago Gomes, and Sandro Pinto are with the ALGORITMI Research Centre/LASI, University of Minho, Braga, Portugal (e-mail: id11207@alunos.uminho.pt; id11674@alunos.uminho.pt; mr.gomes@dei.uminho.pt; sandro.pinto@dei.uminho.pt).}
\thanks{José Martins is with OSYX Technologies, Guimarães, Portugal (e-mail: jose@osyx.tech).}
\thanks{Uwe Moslehner, Kai Dieffenbach, Georg Schmalz, Kajetan Nuernberger, and Thomas Roecker are with Infineon Technologies AG, Munich, Germany (e-mail: Uwe.Moslehner@infineon.com, Kai.Dieffenbach@infineon.com, Georg.Schmalz@infineon.com, Kajetan.Nuernberger@infineon.com, Thomas.Roecker@infineon.com)}
}
%%%%%%%%%%%%%%%%%%%%%%%%%%%%%%%%%%%%

\maketitle
\thispagestyle{plain}
\pagestyle{plain}

%%%%%% -- PAPER CONTENT STARTS-- %%%%%%%%

\begin{abstract}

MCU-based embedded and cyber-physical systems increasingly require protection policies to change at runtime as resources, software, and system composition evolve. Yet widely used COTS MCU protection mechanisms are still largely designed around statically established policies. While they can restrict which domains modify protection primitives, they generally do not constrain what policy an authorized domain may create. As a result, domains may expand protection rules beyond their intended scope or interfere with boundaries established by other domains. We identify \textit{bounded configuration authority} as a missing primitive for secure dynamic policy evolution and introduce an \textit{ownership-aware model} in which explicit ownership and rule containment bound both configuration authority and recursive delegation. Changes to active protection-rule ranges are validated before commit, preserving protection boundaries without extending the normal access-checking datapath. We implement the model in an open-source RISC-V World Checker integrated into a CVA6-based SoC. Our FPGA evaluation shows no degradation in access latency or operating frequency, with 7.1\% LUT and 1.2\% FF overhead for eight rules. In an OTA update case study, direct reconfiguration under delegated authority reduces policy-transition latency by 89.5\% over hypervisor mediation and 94.6\% over trusted firmware mediation, while preserving established authority boundaries. All artifacts will be open-sourced to support the evolution of RISC-V specifications and system-level protection designs.

\end{abstract}

\section{Introduction}
\label{sec:intro}
Modern microcontroller (MCU)-based embedded and cyber-physical systems increasingly consolidate heterogeneous and mixed-criticality workloads over shared memories, peripherals, and accelerators \cite{Gracioli2019,Cerrolaza2020}. Unlike application processors, these systems commonly rely on lightweight, physical-address, region-based protection mechanisms, traditionally configured statically by trusted firmware or a root of trust~\cite{armv8m_memory_model,riscv_security_model,Hardin2018}. Emerging MCU systems, however, increasingly require resource assignments and their associated protection policies to evolve at runtime, driven by changing operating modes and evolving criticality requirements~\cite{Huber2008,Burns2018,ZephyrMemProtection,Danner2014}.

A prominent example is an over-the-air (OTA) software update~\cite{Plappert2023}. A communication domain ($D_C$) may receive an update image into a buffer ($R_b$) and use a hardware accelerator ($H_a$) to process the incoming data. Before installation, both the buffer and accelerator may need to be transferred to an isolated validation domain ($D_V$) for authentication and validation. Copying the image introduces additional memory traffic and latency, while permanent sharing prevents exclusive control. Instead, the protection policy can be updated in place \textcolor{black}{by firmware} to revoke $D_C$'s access and grant access to $D_V$. Identical transitions arise when shared buffers are resized, accelerators are reassigned across operating modes, or software components are incrementally deployed~\cite{Han2005, Schmidt2024}. \textcolor{black}{However, retaining configuration authority in trusted firmware requires mediation for each handover, increasing policy-transition latency.} \textcolor{black}{This dependency also limits autonomy in hierarchical systems, where} trusted boot firmware establishes the protection state and assigns selected resources to a real-time operating system (RTOS) or hypervisor~\cite{Santos2025,Lozano2023}, which may then further partition and assign them among applications or guests~\cite{Pinto2025,Martins2023,Klein2009,ZephyrMemProtection}. \textcolor{black}{Without the ability to delegate configuration authority, these software layers still require firmware intervention to update protection policies.} Direct hardware-supported delegation of configuration authority through these layers avoids repeated privileged software mediation~\cite{Tan2017}, but introduces a fundamental challenge: \textcolor{black}{configuration access alone does not bound policy authority.}

\begin{figure}[t!]
    \centering
    \includegraphics[width=0.88\linewidth]{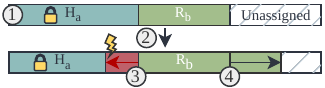}
    \caption{Illustration of unbounded configuration authority over protection rules. $D_C$ in green and $D_V$ in blue. \textcircled{1} initial policy setup, \textcircled{2} configuration delegation, \textcircled{3} malicious overlap by $D_C$, and \textcircled{4} expansion into unassigned space.}
    \label{fig:overlapping_problem}
\end{figure}

%Supporting these scenarios requires more than changing resource-access permissions. In hierarchical systems, i.e., where trusted firmware delegates resource control to a real-time operating system (RTOS) or hypervisor~\cite{Santos2025,Lozano2023}, the authority to configure those permissions may need to be delegated. Trusted boot firmware may establish the protection state and delegate selected resources to these software layers, which may further partition and delegate them among applications or guests~\cite{Pinto2025,Martins2023,Klein2009,ZephyrMemProtection}. Each transition therefore requires privileged software to enforce access-control constraints, validate the resulting policy, and commit the configuration, placing reconfiguration on a trusted software path~\cite{Fiorin2008}. Direct hardware-supported delegation avoids this dependency~\cite{Tan2017}, but introduces a fundamental challenge: configuration access and policy authority are not equivalent.

Protection mechanisms used in mainstream commercial of-the-shelf (COTS) MCUs commonly expose memory-mapped configuration interfaces through which trusted software programs rules associating physical address ranges or resources with execution contexts and access permissions~\cite{ArmTZC380,iopmp}. Access to these interfaces may itself be restricted through additional protection rules~\cite{traveo_t2g_body_um}, execution-domain identifiers~\cite{lpc55_um,sam_l10_l11_ds}, or write-protection mechanisms~\cite{aurix_tc3_um,ra8p1_um}. These mechanisms determine \textit{who} may issue an update, but generally do not constrain \textit{what policy} an authorized domain may create. Consider again the OTA update. The boot process delegates configuration authority over $R_b$ to $D_C$, making $D_C$ the only domain authorized to reprogram the corresponding protection rule. However, $D_C$ may then expand that rule beyond the intended subregion (Figure~\ref{fig:overlapping_problem}). In mechanisms where multiple matching rules contribute to the effective policy, $D_C$ may also create an overlapping rule that weakens protection over a region belonging to another domain. In both cases, $D_C$ modifies only configuration state it controls, yet affects policy outside its intended authority. Existing mechanisms can therefore transfer configuration access without necessarily binding that authority to an enforceable policy boundary.

\textcolor{black}{We identify \textit{bounded configuration authority} as a missing architectural primitive to constrain the policy changes an authorized domain may create. It allows protection policies to evolve at runtime with hardware enforcing \textit{who} may modify each rule and \textit{which} updates are valid.} Moreover, this authority must compose hierarchically: a receiving domain should be able to refine and further delegate its authority without escaping boundaries established by previous owners. Our key observation is that \textit{this hierarchy does not need to be represented through a separate delegation hierarchy}. Explicit rule ownership identifies the domain holding configuration authority, while active enclosing rules naturally encode the bounds within which contained rules may be refined and recursively delegated. \textcolor{black}{In the OTA example, firmware can retain an enclosing rule over $R_b$ and delegate a contained rule to $D_V$, allowing $D_V$ to grant or revoke $D_C$’s access within policy bounds without further firmware mediation.} Changes to protection-rule address ranges are validated against the committed protection state before becoming active. Because configuration changes are infrequent relative to normal resource accesses, this validation can execute on the configuration path without extending the critical access-checking datapath.

Based on these observations, we introduce an ISA-independent \textit{ownership-aware model} \textcolor{black}{for rule-based system-level protection mechanisms. The model defines three architectural semantics: (1) explicit per-rule ownership, (2) protection-state-derived authority bounds, and (3) validate-before-commit reconfiguration.} Ownership determines \textit{who} may modify a rule, while rule containment bounds the policy scope that may be recursively delegated. Invalid updates are rejected without exposing candidate state to the access-checking datapath, enabling direct runtime reconfiguration while preserving protection boundaries established by preceding domains.

We implement the model in an open-source RISC-V World Checker integrated into a CVA6-based system-on-chip (SoC) and evaluate it through FPGA measurements and complete OpenSBI, Bao, and Zephyr deployments. The design preserves baseline access latency while adding 7.1\% LUT and 1.2\% FF overhead for an eight-rule checker. In an OTA update case study, direct delegation reduces policy-transition latency by 89.5\% over hypervisor mediation and 94.6\% over trusted firmware mediation while preserving authority boundaries. Beyond the prototype, the proposed mechanisms are expected to provide an architectural basis for system-level protection checkers in next-generation RISC-V MCUs. We will open-source all artifacts and expect the results to inform ongoing ratification and architectural evolution of system-level specifications within RISC-V International.

In summary, this paper makes the following contributions:

\begin{itemize}
    \item \textcolor{black}{We identify\textit{ bounded configuration authority} as a missing architectural primitive in rule-based system-level protection mechanisms: existing designs can restrict who modifies protection state, but not the policy scope that an authorized configurator may affect (\S~\ref{sec:background}).}

    \item \textcolor{black}{We introduce an \textit{ISA-independent architectural model} that combines explicit per-rule ownership with protection-state-derived authority bounds, enabling configuration authority to be recursively delegated without a separate delegation hierarchy (\S~\ref{sec:model}).}

    \item \textcolor{black}{We design an \textit{off-datapath validation microarchitecture} that checks candidate protection-state updates before commitment, enabling bounded reconfiguration without affecting latency-critical resource-access (\S~\ref{sec:model} and~\ref{sec:obwc}).}

    \item \textcolor{black}{We implement the model in a RISC-V World Checker and evaluate its hardware cost, reconfiguration latency, and scalability, together with complete OpenSBI, Bao, and Zephyr deployments demonstrating direct bounded runtime resource handover (\S~\ref{sec:implementation} and \S~\ref{sec:evaluation}).}

\end{itemize}

\section{Scope, Background, \& Motivation}
\label{sec:background}

MCU-class SoCs differ fundamentally from application processing units (APUs) in how system isolation is enforced. APUs commonly rely on MMUs and page tables for virtual-memory translation, together with IOMMUs~\cite{arm_arch_ref,riscv_iommu,arm_smmu}, often managed by rich OS or hypervisor stacks~\cite{Martins2023}, while MCUs typically use lightweight physical-address region-based mechanisms~\cite{ArmTZC380,riscv_security_model}. \textcolor{black}{CPU-local memory protection units (MPUs) constrain accesses issued by the processor, but do not cover transactions issued independently by DMA engines or accelerators. System-level protection extends enforcement to these initiators through target-side checkers, which enforce access permissions before transactions reach protected memories or peripherals. Protection policies are expressed as address-range rules that associate physical regions with per-domain access permissions. These rules are enforced using transaction security attributes or domain identifiers propagated through the SoC.} This design minimizes area, latency, and software complexity, but offers less flexibility.

%MCU-class SoCs differ fundamentally from application processing units (APU) in how system isolation is enforced. APUs commonly rely on MMUs and page tables for virtual-memory translation, together with IOMMUs~\cite{arm_arch_ref,riscv_iommu,arm_smmu}, all managed by rich OS or hypervisor stacks~\cite{Martins2023}, while MCUs typically use lightweight physical-address region-based mechanisms, such as memory protection units (MPUs) and target-side checkers~\cite{ArmTZC380,riscv_security_model}. Protection policies are expressed as address-range rules and enforced using transaction security attributes or domain identifiers propagated through the SoC. This design minimizes area, latency, and software complexity, but offers less flexibility.

\subsection{Dynamic Policy Evolution}

Dynamic policy evolution involves two distinct dimensions: resource access and configuration authority. In the OTA update workflow introduced earlier, the update buffer and cryptographic accelerator are handed from $D_C$ to $D_V$ by updating their protection policies in place. Unlike copying or permanent sharing, this approach applies uniformly to memories and peripherals while preserving contents or device state and maintaining exclusive access. However, if configuration authority remains with the initial trusted domain, every handover still requires privileged software mediation.

This limitation becomes more relevant as MCU-based systems integrate independently developed components whose resource requirements and access policies may not be fully known to a central authority~\cite{Bai2021,Lee2023}. In domains such as automotive, where subsystems are commonly developed by multiple third parties~\cite{Cheah2016,Lee2023}, configuration authority may therefore need to be delegated across software layers. An initial trusted domain can establish protection boundaries and delegate authority over selected resources to secondary domains, which may in turn refine the corresponding policies and further delegate that authority while remaining within previously established bounds. Resource handover thus determines \textit{who may access} a resource, whereas delegation determines \textit{who may configure} its protection policy. Supporting direct and recursive delegation thus requires hardware to constrain the policy scope available to each receiver.

\subsection{Current Industry Approaches}

\begin{figure}[t!]
    \centering
    \includegraphics[width=\linewidth]{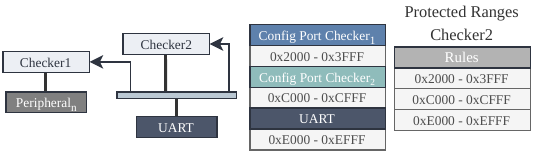}
    \caption{Two instantiations of the rule-based self-protection principle. Checker\textsubscript{2} protects its own configuration interface using one of its own rules. Simultaneously, Checker\textsubscript{2} acts as a centralized configuration authority, mediating write access to Checker\textsubscript{1}'s MMIO interfaces.  }
    \label{fig:overview_approaches}
\end{figure}

To assess how existing platforms support dynamic policy evolution, we survey representative MCU protection mechanisms from major manufacturers. We also consider a rule-based RISC-V configuration, illustrated in Figure~\ref{fig:overview_approaches}, representative of current system-level protection proposals~\cite{worlds,iopmp}. Table~\ref{tab:platform_requirements} compares these platforms across five reconfiguration capabilities. Resource handover (C1) allows access to a resource to be reassigned in place. Configuration-authority transfer (C2) allows another domain to modify the corresponding protection state, while recursive delegation (C3) allows that domain to further transfer this authority. Valid-state commitment (C4) ensures that a new configuration becomes active only after successful validation, and bounded configuration authority (C5) constrains such modifications to the scope established by the delegator.

\begin{table}[t!]
    \caption{Platform support for runtime reconfiguration.
    C1: resource handover; C2: configuration-authority transfer; 
    C3: recursive delegation; C4: valid-state commitment;
    C5: bounded configuration authority.}
    \label{tab:platform_requirements}
    \centering
    \small
    \renewcommand{\arraystretch}{1.15}
    \setlength{\tabcolsep}{3pt}

    \begin{tabular}{
        >{\centering\arraybackslash}p{4.4cm}
        >{\centering\arraybackslash}p{.55cm}
        >{\centering\arraybackslash}p{.55cm}
        >{\centering\arraybackslash}p{.55cm}
        >{\centering\arraybackslash}p{.55cm}
        >{\centering\arraybackslash}p{.55cm}}
        \toprule
        \textbf{Platform}
        & \textbf{C1}
        & \textbf{C2}
        & \textbf{C3}
        & \textbf{C4}
        & \textbf{C5} \\
        \midrule

        Most TZ-M-based~\cite{lpc55_um,stm32h5_rm,stm32l5_rm,stm32u5_rm,sam_l10_l11_ds}
        & \cmark
        & \xmark
        & \xmark
        & \xmark
        & \xmark \\

        \midrule
        \begin{tabular}[c]{@{}c@{}}
        Renesas RA8P1~\cite{ra8p1_um} / RA8M2~\cite{ra8m2_um} \\
        Renesas RH850/U2A~\cite{rh850_um}  \\
        NXP i.MX RT10xx~\cite{imxrt1010_rm} \\
        Infineon Aurix TC3~\cite{aurix_tc3_um}
        \end{tabular}
        & \cmark
        & \halfmark
        & \xmark
        & \xmark
        & \xmark \\

        \midrule
        \begin{tabular}[c]{@{}c@{}}
        Infineon TRAVEO T2G~\cite{traveo_t2g_body_um} \\
        Rule-based RISC-V* \\
        \end{tabular}
        & \cmark
        & \cmark
        & \cmark
        & \xmark
        & \xmark \\

        \midrule
        Infineon Aurix TC4~\cite{aurix_tc4dx_um} 
        & \cmark
        & \cmark
        & \cmark
        & \halfmark
        & \xmark \\
        
        \midrule
        \small \textit{\textbf{RISC-V with OB-WC} }
        & \cmark
        & \cmark
        & \cmark
        & \cmark
        & \cmark \\
        
        \bottomrule
    \end{tabular}

    \vspace{2ex}
    \cmark~Full support,
    \halfmark~Partial support,
    \xmark~No support. \\
    *\textit{Representative setup, elaborated in Figure~\ref{fig:overview_approaches}.}
\end{table}

\mypara{Protecting Configuration State.}
Existing platforms use several mechanisms to determine which domains may modify protection state. Most TZ-M-based systems restrict access to configuration interfaces through Secure and privileged transaction attributes~\cite{lpc55_um,stm32h5_rm,stm32l5_rm,stm32u5_rm,sam_l10_l11_ds}. Other platforms provide programmable filtering mechanisms. For example, the Renesas RA family uses MMPUSARx to specify which security and privilege attributes may modify selected protection state~\cite{ra8m2_um,ra8p1_um}, \textcolor{black}{while Infineon Aurix TC3 uses ACCEN to authorize initiators according to their TAG IDs, with ACCEN itself separately protected by mechanisms such as Safety ENDINIT~\cite{aurix_tc3_um}.} A third approach places one protection mechanism on the configuration path of another, as in Renesas RH850/U2A protection guards~\cite{rh850_um}, Infineon TRAVEO T2G protection pairs~\cite{traveo_t2g_body_um}, and rule-based RISC-V checkers. Infineon Aurix TC4 provides the closest realization of explicit configuration authority by associating protected interfaces with owner domains through PROT~\cite{aurix_tc4dx_um}, allowing the owner to transfer that authority directly to another domain.

\mypara{Transferring Configuration Authority.}
All surveyed platforms support resource handover by updating access permissions, although most TZ-M-based systems require secure world mediation. Renesas RA, NXP i.MX RT10xx~\cite{imxrt1010_rm}, Renesas RH850/U2A, and Infineon Aurix TC3 allow selected software domains or initiators to modify protection state, but this authority is constrained by privilege, boot-time configuration, or limited granularity over the protected interface. \textcolor{black}{They therefore provide only partial configuration-authority transfer: a receiving domain may modify protection state without being able to change which domains have that permission. This prevents recursive delegation.} Infineon Aurix TC4 explicitly supports ownership transfer, while Infineon TRAVEO T2G and rule-based RISC-V configurations can reassign authority by modifying the protection state governing access to configuration interfaces. In these cases, the receiver can subsequently transfer that authority again.

\mypara{Unbounded Configuration Authority.}
Despite supporting several of these capabilities, existing mechanisms primarily constrain \textit{who} may modify protection state, not \textit{what policy} an authorized domain may establish. Aurix TC4 provides a representative example of this distinction: PROT identifies which domain controls an interface, but does not bound the protection policy that the receiving domain may subsequently create. \textcolor{black}{An authorized domain may therefore expand a rule beyond its intended allocation or, where any matching rule can grant access (accretive permissions), create an overlapping rule that bypasses protection established by another rule.} More generally, the surveyed mechanisms can transfer configuration authority without binding it to an enforceable policy scope. Thus, while existing MCUs provide several building blocks for dynamic reconfiguration, none combines recursive delegation, bounded configuration authority, and valid-state commitment.

\subsection{Architectural Observations}

The surveyed platforms yield a common architectural insight. Existing mechanisms support resource handover and, in some cases, recursive delegation, but represent configuration authority primarily through access to protection state. This determines \textit{who} may issue an update, but not the policy scope that the update may affect. Together with the use cases, this insight leads to four architectural observations.

\mypara{Observation 1: Configuration Access Does Not Bound Policy Authority.}
Safe dynamic reconfiguration requires configuration authority to be both explicit and bounded. \textcolor{black}{A domain may modify only a rule it controls and still affect resources outside its allocation by expanding that rule’s range.} Therefore, hardware must identify not only which domain may modify a protection object, but also the policy scope within which that authority may be exercised.

\mypara{Observation 2: Software Hierarchies Require Composable Delegation.}
System software is inherently hierarchical. Trusted firmware may delegate responsibility to an OS or hypervisor, which may further divide it among applications or guests. Configuration authority must compose across the same layers: each receiver should be able to refine and further delegate its assigned authority while remaining within boundaries established by preceding domains. Bounds introduced at one layer must therefore remain effective across all subsequent delegation levels.

\mypara{Observation 3: Rule Containment Can Encode Delegation Boundaries.}
The hierarchy required for bounded delegation can be represented directly by the protection rules themselves. A domain that owns an enclosing active rule can define the configuration boundary within which another domain controls a contained rule. Once ownership of the contained rule is transferred, the receiver may refine and further delegate protection within that range, while the enclosing rule preserves the boundary established by the previous owner. Ownership and range containment can therefore jointly encode hierarchical delegation without explicit parent identifiers or a separate capability tree.

\mypara{Observation 4: Configuration Validation Can Execute Off the Datapath.}
Normal resource accesses require low-latency enforcement, whereas configuration updates are comparatively infrequent. Policy validation can therefore execute over multiple cycles on the configuration path without extending the critical access-checking datapath. A candidate update can be staged and checked against the active rules while the committed policy continues to enforce accesses. Valid candidates are committed after validation, while rejected candidates leave the active policy unchanged.

\section{Bounded Configuration Authority}
\label{sec:model}

Building on Section \ref{sec:background}, we introduce a bounded configuration-authority model combining explicit rule ownership, protection-state-derived authority bounds, and validate-before-commit updates. Each rule becomes an independently owned configuration object, separating configuration authority from resource access. Ownership can be transferred recursively, while active enclosing rules bound how that authority may be refined and further delegated. Thus, ownership determines \textit{who} may update a rule, containment bounds \textit{what policy} it may create, and valid-state commitment determines \textit{which configuration} may become active.

\subsection{System Model and Assumptions}

Despite implementation differences, the surveyed platforms share a common structure: hardware-propagated transaction initiator identifiers are evaluated by target-side protection mechanisms configured through memory-mapped interfaces. We adopt this structure as the basis for our model.

\mypara{Domains and Access Enforcement.}
We consider an MCU partitioned into multiple execution domains, each associated with a system-level identifier. A domain is an abstract encapsulated entity whose granularity depends on the software stack: it may represent (1) a hypervisor VM / an RTOS instance managed by a TEE or simply (2) an RTOS task / OS process. Examples of system-level identifiers include a TRAVEO T2G~\cite{traveo_t2g_body_um} protection context or a RISC-V World Identifier (WID)~\cite{worlds}. Every transaction carries the identifier of its issuing domain, together with the target address and requested operation. These identifiers are assigned and propagated by hardware and cannot be forged by the initiator. Target-side checkers mediate accesses to memories, peripherals, and other memory-mapped resources from both CPU and non-CPU initiators (e.g., DMAs). Access policies are expressed through rules specifying an address range, the domains and corresponding permissions (i.e., read, write, execute) within that range, and whether the rule is enabled.

\mypara{Configuration Model.}
Target-side checkers' rules are exposed through memory-mapped configuration registers. Each rule comprises an address range, resource-access permissions, configuration state, and ownership state. The first three determine how the rule contributes to the effective resource-access policy, while ownership identifies the domain authorized to modify it, independently of the domains allowed to access the protected resource. Configuration writes carry the requesting domain identifier, allowing the checker to authenticate the issuer. Each rule therefore serves both as part of the resource-access policy and as a protection object over which configuration authority may be assigned.

\mypara{Boundary Assumptions.}
The trusted root initially establishes ownership boundaries over every address range whose subsequent allocation must remain constrained. Address space not covered by an active rule has no established delegation boundary and may be claimed by any domain, subject to the validation semantics defined later. Delegation is hierarchical: a domain retaining an enclosing rule remains trusted with respect to policies configured within that range.

\mypara{Threat Model.}
A trusted root establishes the initial protection state and configuration authority at boot. Other domains may subsequently receive authority over selected protection objects and may be compromised or malicious, abusing legitimate configuration access to expand policies beyond their assigned scope or interfere with protections established by earlier domains. We trust the protection hardware and initial configuration. Identifier forgery, checker bypass, physical attacks, and denial-of-service attacks are outside our scope. The model is ISA-independent and applies to rule-based target-side protection mechanisms supplied with trusted transaction metadata identifying the issuing domain.

\subsection{Explicit Configuration Ownership}
\label{sec:3_2}

We make configuration authority explicit by associating each rule with an owner identifier, independently of its resource-access permissions. A domain may therefore own a rule without accessing the protected region, while access does not confer reconfiguration authority. Configuration writes are authenticated using the requesting domain's identifier and follow three ownership rules:

\mypara{1.} Only the owner of an owned rule may modify its configuration. Requests from other domains are rejected.

\mypara{2.} Ownership persists while a rule is disabled. Disabling removes it from the data-access policy but does not release ownership or allow others to reconfigure it.

\mypara{3.} An ownerless rule cannot be enabled. An ownerless rule may be claimed by any domain, and the trusted root need not pre-assign ownership to every rule. Preparatory writes to ownerless rules are permitted, but are not protected against concurrent modification; software requiring integrity during preparation must therefore claim the rule first.

\begin{figure}[t!]
    \centering
    \includegraphics[width=.7\linewidth]{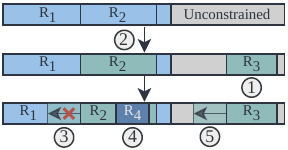}
    \caption{Bounded configuration authority. Each color identifies a different execution domain: ~\textcircled{1} ownership acquisition of an ownerless rule, ~\textcircled{2} and~\textcircled{4} contained-rule delegation to a new domain, ~\textcircled{3} rejected expansion across another owner’s boundary, and ~\textcircled{5} valid expansion into unconstrained space.}
    \label{fig:supported_operations}
\end{figure}

Ownership is established by claiming an ownerless rule, as illustrated by action~\textcircled{1} in Figure~\ref{fig:supported_operations}. The claiming domain becomes its sole configuration authority without automatically gaining access to the protected resource. An owner may later transfer the rule to another domain~\textcircled{2}; only ownership changes, while the range, enable state, and access permissions remain unchanged. After the transfer, the previous owner can no longer modify the rule.

Delegation becomes recursive when a receiver uses its authority to establish another delegation level. In action~\textcircled{2}, a domain retains ownership of an enclosing rule while transferring a contained rule to another domain. The enclosing rule preserves the delegator's authority boundary, while the receiver may refine or further delegate protection within that boundary~\textcircled{4}. Repeating this operation creates nested configuration authority, with each delegator retaining an enclosing rule and each receiver controlling a contained rule. Ownership and range containment therefore encode hierarchical delegation without requiring explicit parent identifiers or exposing the checker's entire configuration interface.

\subsection{Bounded Policy Validation}

Ownership determines whether a domain may issue an update, but authorized changes may still violate existing boundaries or expose inconsistent state. We therefore enforce three validation and commitment rules:

\mypara{1.} Any change to a rule's active address range must be validated before affecting access checks. This includes enabling a rule or modifying an enabled range. Disabled rules may be prepared without validation, while independent permission or ownership changes require none because they preserve the active range.

\mypara{2.} \textcolor{black}{A candidate may overlap another domain’s enabled rule only if its entire address range is covered by enabled, committed rules owned by the requester. Coverage may be provided by a single rule or by the union of multiple rules.} Additionally, for an update on an enabled rule, its currently committed range also provides such a boundary. Thus, action~\textcircled{3} in Figure~\ref{fig:supported_operations} is rejected for crossing another owner's boundary, while action~\textcircled{4} remains within a requester-owned contained region. Expansion into unconstrained space is also allowed when no ownership boundary is crossed, as in action~\textcircled{5}.

\mypara{3.} Only a complete and valid candidate may replace the active rule. Acceptance atomically commits the new state, while rejection leaves the previous state unchanged. Atomicity applies per request; updates spanning multiple requests must be ordered by software so that every intermediate committed state satisfies the intended policy.

These semantics derive delegation boundaries directly from active protection state, avoiding separate metadata. The trusted root must therefore cover any address space whose future allocation must remain constrained; other space remains available. This yields a directional guarantee: descendants cannot escape ancestor-established boundaries or affect policy beyond them, while ancestors retaining enclosing rules remain authoritative over nested regions.

\textcolor{black}{Controlled overlap supports nested delegation by allowing a receiver’s rule to coexist with the active enclosing rule retained by its delegator, while validation constrains updates to the requester’s previously established authority.} It can additionally reduce the number of rules required to represent delegated policies. With strictly disjoint ranges, delegating a subregion may require partitioning the enclosing range around it; retaining the enclosing rule and adding a contained rule avoids this partitioning and allows the same boundary to support multiple regions belonging to different domains.

\section{Ownership-Bounded World Checker}
\label{sec:obwc}

\textcolor{black}{To explore the proposed semantics, we developed a low-cost microarchitectural extension to a RISC-V World Checker (WC), whose rule-based, target-side organization matches the model of Section~\ref{sec:model}. We call this design the \textit{Ownership-Bounded World Checker} (OB-WC). RISC-V is a suitable target because World Checkers and IOPMP resemble the rule-based MCU mechanisms surveyed in Section~\ref{sec:background}.}

% We realize the proposed semantics through a low-cost microarchitectural extension to a RISC-V World Checker (WC), whose rule-based, target-side organization matches Section~\ref{sec:model}'s model. We call this design the \textit{Ownership-Bounded World Checker} (OB-WC). RISC-V is a suitable target because World Checkers and IOPMP follow a rule-based organization similar to the MCU mechanisms surveyed in Section~\ref{sec:background}. % and are currently under active standardization.

\begin{figure}[t!]
    \centering
    \includegraphics[width=\linewidth]{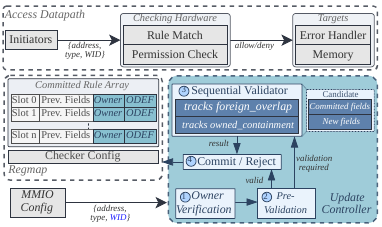}
    \caption{OB-WC architecture. Components highlighted in blue are added by OB-WC. Configuration requests pass through the update controller, which performs \textcircled{1} owner verification, \textcircled{2} pre-validation, \textcircled{3} sequential validation against the committed rule array when required, and \textcircled{4} commit or rejection of the candidate state. Ownership metadata is stored alongside each rule, while the baseline access datapath remains unchanged.
    }
    \label{fig:obwc_arch}
\end{figure}

OB-WC adds two mechanisms to the baseline WC: explicit ownership metadata for each rule and an update controller on the memory-mapped configuration path. The controller authenticates configuration requests, validates candidate ranges when required, and commits only accepted updates. The committed rules continue to drive the resource access-checking datapath while candidate configurations are validated independently over multiple cycles. Figure~\ref{fig:obwc_arch} summarizes this architecture, separating the unchanged access datapath from the ownership-aware configuration and validation path.

\subsection{Slot and Configuration-Path Extensions}

In the baseline WC, each protection rule is represented by a slot comprising an address register, a configuration register, and a permissions register. The address register defines the corresponding protected range, the configuration register holds the slot state, and the permissions register encodes per-World access permissions.

OB-WC extends each slot with two ownership fields: \texttt{OWNER}, which identifies the World holding configuration authority, and \texttt{ODEF}, which indicates whether ownership is defined. These fields affect only configuration authority and do not contribute to the resource-access policy. When \texttt{ODEF} is set, writes to the slot are authorized by comparing the WID carried by the request against the committed \texttt{OWNER}. When \texttt{ODEF} is clear, the slot is ownerless and must remain disabled. Any World may prepare an ownerless slot, claim it, or assign it directly to another World; once ownership is defined, only the committed owner may modify or transfer the slot.

An update controller is inserted between the WC configuration interface and the slot registers. For each write, it receives the requester WID, target slot, and proposed register value, and constructs the corresponding candidate slot state from the proposed and committed fields. Since a single address-register write encodes the complete range, no software-visible shadow registers are required. The committed state is active throughout validation: accepted candidates replace it atomically at the granularity of the configuration request, while rejected requests leave it unchanged.

\subsection{Configuration Update Flow}

Configuration requests proceed through four stages: owner verification, pre-validation, optional range validation, and final commit or rejection. Configuration writes are serialized locally, allowing the controller to maintain a single candidate state and preventing concurrent writes from affecting validation. Atomicity applies to an individual configuration request; updates spanning multiple slot registers or multiple slots must be ordered by software so that every intermediate committed state preserves the intended protection policy.

First, the requester is authorized against the committed ownership state of the target slot. If \texttt{ODEF} is set, the requester WID must match \texttt{OWNER}; ownerless slots may be prepared while disabled. Pre-validation then determines whether range validation is required. Permission changes, ownership transfers, rule disablement, and modifications to disabled slots preserve the active range and can therefore commit without scanning the rule array. Validation is also bypassed when an enabled rule is refined to a range fully contained within its currently committed range. Requests that enable, expand, or relocate an active range instead proceed to range validation.

A request may also activate a slot while transferring its ownership. Authorization and validation use the committed owner, since that domain is establishing the new active rule and delegating its configuration authority. If accepted, the candidate owner becomes active only when the complete candidate state is committed. For ownerless slots, the requesting World provides the authority used during validation. While validation is in progress, further writes to the slot registers are stalled. This simplifies the controller design, although pending writes could instead be buffered in a small FIFO. Since rejection produces no separate software-visible indication, software must read back the target register to determine whether the update was committed.

\subsection{Sequential Range Validation}

Sequential range validation determines whether a candidate remains within the configuration authority encoded by the committed rules. The controller scans the enabled slots, processing one active slot per cycle; disabled slots are skipped. Validation latency thus depends on the number of enabled rules rather than the total number of implemented slots. 

During the scan, the controller tracks two conditions: whether the candidate overlaps a rule not owned by the requester and whether an enabled requester-owned rule fully contains the candidate. The candidate is accepted if $\neg\textit{foreign\_overlap} \lor \textit{owned\_containment}$. $\textit{foreign\_overlap}$ indicates that the candidate intersects an enabled rule owned by another domain, while $\textit{owned\_containment}$ establishes that the complete candidate remains within an active boundary already controlled by the requester. Therefore, a candidate with no cross-owner overlap can be accepted directly, whereas a candidate overlapping another owner's rule is accepted only if an enclosing requester-owned rule contains it completely. The scan continues after detecting a cross-owner overlap because a qualifying enclosing rule may appear in a later slot. Conversely, once such a containing rule is found, the candidate can be accepted immediately: full containment establishes that all candidate effects remain within a boundary already controlled by the requester. \textcolor{black}{Checking containment using a single rule simplifies the sequential validator while satisfying the model’s coverage condition. However, merging separately delegated regions may require retaining a requester-owned rule that encloses their combined range, rather than relying on the combined coverage of the individual rules.}

Bounded configuration authority does not change how the underlying WC combines data-access permissions across overlapping rules.
Software must therefore configure rule permissions so that their combined effect enforces the intended resource-access policy.

\subsection{Integration and Datapath Implications}

The OB-WC extensions are confined to the configuration path. Normal resource accesses continue to use the baseline WC logic and the committed address ranges and permissions. Ownership metadata, candidate state, and validation results do not participate in access checks, and the update controller introduces no additional stage in the access-checking datapath. While a candidate is being validated, the committed policy remains active and normal accesses proceed uninterrupted. Once accepted, the candidate replaces the committed state at the single-clock-cycle granularity, so accesses observe either the previous or updated policy without exposing an intermediate configuration. \textcolor{black}{Transactions allowed under the previous policy may still be in progress when a new policy becomes active. Before handing over a resource, software must therefore stop the relevant initiators from issuing new transactions and wait for outstanding transactions and ongoing peripheral operations to complete. OB-WC validates configuration changes but leaves synchronization of resource accesses to software.}

Integration requires only the requester WID already propagated with WC transactions to be provided to the update controller together with the target register and proposed value. No additional identity mechanism or explicit delegation-tree interface is required. Configuration writes are serialized locally by each checker, but this does not stall its access-checking path. However, if configuration and resource traffic share an upstream interconnect, both may still experience contention from the surrounding arbitration logic. OB-WC therefore preserves the normal access datapath while shifting the additional cost of bounded reconfiguration to the comparatively infrequent configuration path.

\begin{figure}[t!]
    \centering
    \includegraphics[width=0.85\linewidth]{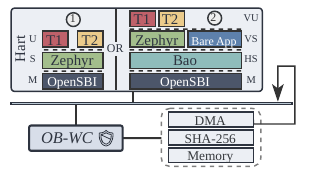}
    \caption{FPGA prototype and implemented software configurations. Each software color represents a different World. (1) OpenSBI+Zephyr, (2) OpenSBI+Bao+Zephyr and a bare-metal guest. A single OB-WC protects memory and the MMIO interfaces of the SHA-256 accelerator and DMA.}
    \label{fig:implementation}
\end{figure}

\begin{figure*}[t!]
    \centering
    % ---------- Row 1: Rules ----------
    \subfloat[LUT comparison between WC, RP-WC, ORP-WC, and OB-WC.]{
        \includegraphics[width=0.48\textwidth]{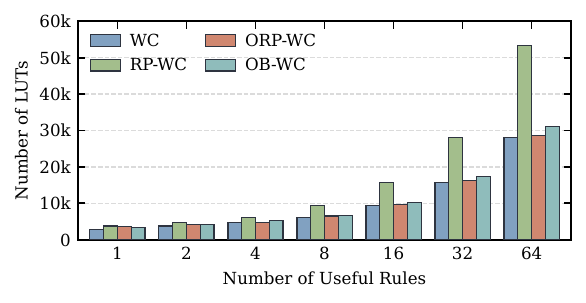}
        \label{fig:lut_chart}
    }
    \hfill
    \subfloat[FF comparison between WC, RP-WC, ORP-WC, and OB-WC.]{
        \includegraphics[width=0.48\textwidth]{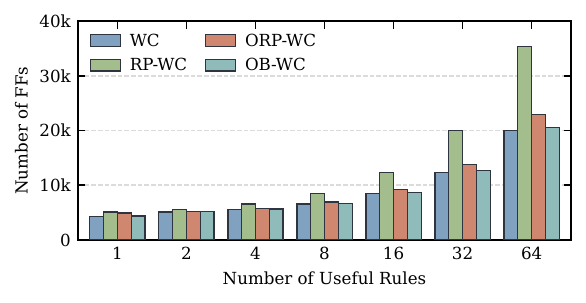}
        \label{fig:ff_chart}
    }

    \caption{Resource utilization comparison between WC, RP-WC, ORP-WC, and OB-WC as the number of useful rules increases.}
    \label{fig:wgc_resource_comparison}
\end{figure*}

\section{Implementation}
\label{sec:implementation}

We implement OB-WC in a single-hart CVA6-based~\cite{zaruba2019cost} MCU configuration (SPMP instead of MMU~\cite{Rodriguez2026SPMP}), and deploy the prototype on a Xilinx Genesys2 FPGA. We extend CVA6 with architectural state to identify the active RISC-V World and tag outgoing transactions with its WID. WIDs propagate through an AXI interconnect to the protection mechanisms, while enforcement remains confined to the target-side OB-WC. This checker protects a memory region and the memory-mapped SHA-256 accelerator and DMA engine. The DMA also acts as an interconnect initiator: it records the WID of the World configuring it and attaches that identifier to its generated transactions, ensuring that non-CPU accesses traverse the same protection mechanism. The prototype supports eight Worlds and an eight-slot OB-WC.

\mypara{Software Support.}
We integrate OB-WC across OpenSBI~\cite{opensbi}, Bao~\cite{Martins2020}, and Zephyr~\cite{zephyr}, allowing configuration authority at multiple software layers. This software stack has served as a reference configuration in the ratification of RISC-V MCU specifications, notably SPMP~\cite{Rodriguez2026SPMP}. OpenSBI establishes the WIDs used by higher layers, initializes protection state, and provides OB-WC configuration routines. Bao assigns a distinct World to each guest, updates the active WID on guest switches, establishes OB-WC rules, and may delegate ownership of selected rules to its guests. Zephyr runs its kernel in a dedicated World while assigning distinct Worlds to individual tasks; its scheduler updates the active WID during task switches, and an OB-WC driver allows tasks or the kernel to modify rules under delegated authority. We exercise the prototype with two complete software configurations: OpenSBI with Zephyr, and OpenSBI with Bao hosting Zephyr and a bare-metal guest. Figure~\ref{fig:implementation} summarizes the software hierarchy and protected hardware used in both configurations.

\section{Evaluation}
\label{sec:evaluation}

We evaluate OB-WC along five dimensions: hardware cost, datapath impact, reconfiguration latency, system-level utility, and security enforcement. These objectives motivate five research questions:

\mypara{RQ1: Hardware Cost.}
What area and storage overheads are introduced by bounded configuration authority?

\mypara{RQ2: Datapath Cost.}
Does bounded-authority reconfiguration affect normal access latency or the maximum operating frequency of the checker?

\mypara{RQ3: Reconfiguration Cost.}
What latency does bounded reconfiguration introduce compared with unprotected and rule-protected configuration paths?

\mypara{RQ4: System Utility.}
Can OB-WC support hierarchical delegation of configuration authority followed by direct runtime resource handovers, and what is the benefit relative to continued software mediation?

\mypara{RQ5: Security Enforcement.}
Does OB-WC prevent delegated domains from modifying protection state outside their assigned authority while preserving committed state after rejected updates?

\subsection{Hardware Cost} 

To answer RQ1, we synthesize four WC configurations for the Genesys2 FPGA. The baseline WC provides no dedicated configuration protection. \textcolor{black}{The rule-protected WC (RP-WC) pairs each useful rule with a second rule protecting both rules’ configuration registers. Configuration accesses pass through the checker’s existing checking interface, without dedicated internal routing. The optimized RP-WC (ORP-WC) introduces an internal address-based switch to select between useful and protecting rules and uses fixed ranges and limited configuration for the protecting rules.} OB-WC instead adds explicit ownership metadata and the update controller described in Section~\ref{sec:obwc}. RP-WC and ORP-WC constrain configuration access but do not provide bounded configuration authority.

For a fair comparison, we normalize all designs by the number of \textit{useful rules}, i.e., rules available for protecting system resources. WC and OB-WC require $N$ physical rules for $N$ useful rules, whereas RP-WC and ORP-WC require $2N$ physical rules since each useful rule requires a protecting rule. We evaluate configurations from one to 64 useful rules (up to 128 physical rules for the rule-protected designs).

Figure~\ref{fig:wgc_resource_comparison} reports LUT and FF utilization. For the eight-rule design, WC requires 6,157 LUTs and 6,537 FFs, while OB-WC requires 6,595 LUTs and 6,617 FFs, corresponding to 7.1\% LUT and 1.2\% FF overhead. At 64 rules, the overhead reaches 10.9\% and 3.0\%, respectively. The FF increase is mainly due to the additional OWNER and ODEF metadata, while the LUT cost stems from the additional update controller and validation logic.

Rule-based self-protection scales less favorably because it duplicates the rule array. RP-WC introduces 52.3\% LUT and 29.5\% FF overhead at eight useful rules, increasing to 90.6\% and 77.0\% at 64. ORP-WC reduces this cost and, at 64 useful rules, requires 8.6\% fewer LUTs but 10.3\% more FFs than OB-WC. Unlike both alternatives, OB-WC preserves all implemented rules for system-resource protection while providing explicit ownership, bounded authority, and validate-before-commit semantics.

\newpage
\mypara{Answer to RQ1.}
OB-WC adds 7.1\% LUT and 1.2\% FF overhead for eight rules, increasing to 10.9\% and 3.0\% at 64, while preserving all rules for system-resource protection.

\subsection{Datapath Cost}

To answer RQ2, we compare access latency and maximum operating frequency between the eight-rule WC and OB-WC implementations. A DMA initiator issues single-beat requests with outstanding and out-of-order transactions disabled, exposing the latency of each individual access. Latency is measured from request issuance until the corresponding response becomes available at the DMA.

Both WC and OB-WC complete each access in two cycles. This follows directly from the architecture: OWNER, ODEF, and the update controller are confined to the configuration path, while normal accesses use the same committed rule state and checking datapath. \textcolor{black}{The frequency comparison additionally checks whether the added configuration logic and connections to the rule array introduce a timing bottleneck during implementation, despite leaving the access-checking logic unchanged.} Under the same implementation flow, WC and OB-WC reach approximately 151\,MHz and 153\,MHz, respectively. The 1.3\% difference is within implementation variation; in both designs, the critical path remains in the existing rule-analysis datapath rather than the added configuration logic.

\mypara{Answer to RQ2.}
OB-WC preserves the two-cycle access latency and checker frequency, keeping bounded reconfiguration off the normal access datapath.

\begin{table}[t!]
	\caption{Configuration latency of the evaluated mechanisms.}
	\label{tab:config_latency}
	\centering
	\scalebox{1}{%
	\begin{tabular}{
	   >{\centering\arraybackslash}p{2.5cm}
	   >{\centering\arraybackslash}p{1.5cm}
	   >{\centering\arraybackslash}p{1.7cm}
	   >{\centering\arraybackslash}p{1.1cm}} 
	
		\toprule
        \parbox{2.5cm}{\centering \small \textbf{Operation}}  
        & \parbox{1.5cm}{\centering \small \textbf{WC w/o protection}}
        & \parbox{1.7cm}{\centering \small \textbf{Rule-based protection}}
        & \parbox{1.1cm}{\centering \small \textbf{OB-WC}} \\
        
        \cmidrule(l){1-4}
        \footnotesize Activate rule / Change active range  
        & \footnotesize 8 cycles 
        & \footnotesize 11 cycles 
        & \footnotesize $9 + N_e$ cycles \\

        \cmidrule(l){1-4}
        \footnotesize Change owner  
        & \footnotesize 8 cycles 
        & \footnotesize 11 cycles 
        & \footnotesize 9 cycles \\

        \cmidrule(l){1-4}
        \footnotesize Change permissions  
        & \footnotesize 8 cycles 
        & \footnotesize 11 cycles 
        & \footnotesize 9 cycles \\
        
	\bottomrule
	\end{tabular}
	}
    \vspace{0.5ex}
    
    \footnotesize $N_e$ denotes the number of enabled rules checked during range validation.
\end{table}

\subsection{Reconfiguration Cost}

To answer RQ3, we use a bare-metal microbenchmark issuing isolated writes to the checker configuration interface. Latency is measured from acceptance of a configuration transaction at the checker port until the corresponding rule state is committed and becomes active, isolating the intrinsic cost of the reconfiguration mechanism.

Table~\ref{tab:config_latency} summarizes the measured latency. WC provides the baseline because configuration writes are applied directly. RP-WC adds three cycles because each request first traverses the rule protecting the configuration interface. OB-WC adds only one cycle for permission and ownership updates that do not require range validation. Operations that enable a rule or modify an active range incur variable latency because the candidate must be checked against the enabled rules. Sequential validation requires $9 + N_e$ cycles, where $N_e$ is the number of enabled rules examined. Disabled rules are skipped and add no validation cycles. \textcolor{black}{For the eight-slot prototype, the scan therefore contributes at most eight cycles, giving a maximum configuration latency of 17 cycles. Finding a requester-owned enclosing rule permits early acceptance, potentially reducing the number of rules examined.} This variable component results from the chosen sequential validator rather than from the ownership semantics; a parallel implementation could trade additional hardware for lower latency.

\mypara{Answer to RQ3.}
OB-WC requires 9 cycles for permission and ownership updates, while range-changing operations require $9 + N_e$ cycles, with variable cost incurred only when range validation is necessary.

\begin{figure}[t!]
    \centering
    \includegraphics[width=0.7\linewidth]{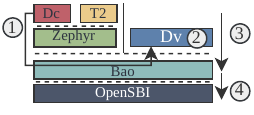}
    \caption{OTA case-study workflow and evaluated configuration paths. The communication task $D_C$ requests validation from $D_V$ through Bao's shared-memory IPC~\textcircled{1}. $D_V$ then reconfigures the resource-protection rules directly~\textcircled{2}, through Bao~\textcircled{3}, or through Bao and OpenSBI~\textcircled{4}.}
    \label{fig:ota_flow}
\end{figure}

\subsection{OTA Update Case Study}

To answer RQ4, we evaluate OB-WC using the OTA workflow introduced in Section~\ref{sec:intro}. The FPGA prototype exercises delegation across OpenSBI, Bao, and isolated application domains, while evaluating the same runtime resource handover under three placements of configuration authority. Figure~\ref{fig:ota_flow} summarizes the software hierarchy and evaluated paths. The case study exercises two distinct operations. First, OpenSBI recursively delegates configuration authority through Bao to $D_V$. Second, $D_V$ uses that authority to hand the update buffer and SHA-256 accelerator between itself and $D_C$. The first changes \textit{who may configure} the protection rules; the second changes \textit{who may access} the protected resources.

\mypara{OTA Workflow.}
Bao hosts a Zephyr guest and an isolated bare-metal $D_V$, while $D_C$ executes as a Zephyr task with a distinct WID. Separate OB-WC rules protect a 16\,KiB update buffer and the MMIO interface of a SHA-256 accelerator. At boot, ownership of both rules is delegated from OpenSBI to Bao and then to $D_V$. The prototype uses a copy of the $D_V$ bare-metal binary as the update image, abstracting only the external transport stage while exercising the resource handover, isolated hashing, and restoration stages in full.
During reception, $D_V$ retains ownership while granting $D_C$ access to the buffer and accelerator. After $D_C$ loads the image and notifies $D_V$ through Bao's shared-memory IPC, $D_V$ revokes $D_C$'s access privileges, computes the digest in the isolated environment, and restores access after validation. Configuration authority remains with $D_V$ throughout the whole flow.

\mypara{Evaluated Configuration Paths.}
We execute the same workflow under three placements of configuration authority. In the direct path~\textcircled{2}, $D_V$ owns both rules and updates their permissions directly. In the Bao-mediated path~\textcircled{3}, Bao retains ownership and applies the changes on behalf of $D_V$. In the OpenSBI-mediated path~\textcircled{4}, OpenSBI retains ownership, requiring each request to traverse Bao and firmware. All three paths apply the same permission changes and establish the same final protection state. \textcolor{black}{Since the ranges are already active, these permission updates bypass sequential range validation. The latency comparison therefore captures software execution and mediation around the same fixed-cost hardware operation.}

\mypara{Mediation Overhead.}
We measure policy-transition latency using a shared hardware cycle counter, from the instant $D_V$ initiates the transition until both permission updates have committed. The measurement includes mediation through Bao or OpenSBI when present, but excludes the preceding notification from $D_C$ and the subsequent hashing operation.
As Table~\ref{tab:latency_mediation} summarizes, direct reconfiguration requires an average of 876 cycles. Bao mediation increases this to 8381 cycles, while OpenSBI mediation requires 16304 cycles. Since all three paths invoke the same application-level operations and establish the same final permissions, the difference isolates the cost of successive software mediation layers. Direct delegation thus reduces policy-transition latency by 89.5\% and 94.6\% relative to Bao and OpenSBI mediation, respectively.

\begin{table}[t!]
	\caption{Reconfiguration latency for different mediation paths in clock cycles.}
	\label{tab:latency_mediation}
	\centering
	\scalebox{1}{%
	\begin{tabular}{
	   >{\centering\arraybackslash}p{2cm} 
	   >{\centering\arraybackslash}p{1.3cm}
	   >{\centering\arraybackslash}p{1.3cm}
	   >{\centering\arraybackslash}p{1.3cm}} 
	
		\toprule
        \parbox{2cm}{\centering \small \textbf{Configuration authority}}  
        & \parbox{1.3cm}{\centering \small \textbf{Min}}
        & \parbox{1.3cm}{\centering \small \textbf{Mean}}
        & \parbox{1.3cm}{\centering \small \textbf{Max}} \\
        
        \cmidrule(l){1-4}
        \small $D_V$ &   \small 871  & \small 876 & \small 915 \\
        
        \cmidrule(l){1-4}
        \small Bao &   \small 8241 & \small 8381 & \small 8419 \\

        \cmidrule(l){1-4}
        \small OpenSBI &   \small 16019 & \small 16304 & \small 16371 \\
		
	\bottomrule
	\end{tabular}
	}
\end{table}

\mypara{Answer to RQ4.}
OB-WC supports recursive delegation of configuration authority across OpenSBI, Bao, and $D_V$. Once $D_V$ becomes the rule owner, it can perform the runtime resource handovers directly, reducing policy-transition latency by 89.5\% and 94.6\% relative to Bao and OpenSBI mediation.

\subsection{Security Enforcement}

Direct delegation is useful only if delegated owners remain confined to the authority established by preceding software layers. To answer RQ5, we exercise OB-WC with valid and adversarial updates derived from the OTA configuration. $D_V$ acts as the delegated owner of the rules protecting the update buffer and SHA-256 accelerator, while surrounding protection state defines the boundaries that its authority must preserve. The tests cover owner-authorized updates, valid range refinement, modification by non-owners, expansion beyond delegated boundaries, unsafe cross-owner overlap, simultaneous activation and ownership transfer outside an established boundary, preservation of committed state after rejection, and access enforcement following reconfiguration. For each request, we record whether it is accepted or rejected and verify both the committed rule state and effective resource-access permissions. \textcolor{black}{The test software and protection configurations will be released to enable reproduction of this setup and comparison with the outcomes reported in Table IV.}

Table~\ref{tab:obwc_security_validation} summarizes the results. Legitimate permission updates and range refinements within delegated boundaries are accepted and committed. Ownership verification rejects modifications by non-owners, while range validation rejects expansions or overlaps that escape the requester's previously established authority. Rejected updates leave the committed rule state unchanged. After $D_C$'s permissions are revoked, its accesses are not allowed while $D_V$ retains access during isolated validation.

\mypara{Answer to RQ5.}
OB-WC permits authorized refinement while preventing non-owners and delegated owners from modifying policy outside their assigned authority, and rejected requests preserve the previously committed state.

\begin{table}[t!]
    \caption{Directed security validation of OB-WC.}
    \label{tab:obwc_security_validation}
    \centering
    \footnotesize
    \renewcommand{\arraystretch}{1.15}
    \begin{tabularx}{\columnwidth}{
        >{\raggedright\arraybackslash}X
        >{\centering\arraybackslash}p{2.4cm}}
        \hline
        \textbf{Tested Condition} 
        & \textbf{Observed Outcome} \\
        \hline

        \multirow{2}{*}{Owner-authorized permission update}
        & Accepted and committed \\
        \hline

        \multirow{2}{*}{Range refinement within delegated boundary}
        & Accepted and committed \\
        \hline

        \multirow{2}{*}{Rule modification by a non-owner}
        & Rejected, state unchanged \\
        \hline

        \multirow{2}{*}{Range expansion beyond delegated boundary}
        & Rejected, state unchanged \\
        \hline

        Cross-owner overlap without a requester-owned enclosing boundary
        & Rejected, state unchanged \\
        \hline

        Enable rule with simultaneous ownership change outside delegated boundary
        & Rejected, state unchanged \\
        \hline

        $D_C$ access after permission revocation
        & Denied \\
        \hline

        $D_V$ access during isolated validation
        & Allowed \\
        \hline
    \end{tabularx}
\end{table}

\section{Discussion}

\mypara{Explicit Ownership as an Architectural Primitive.}
Existing platforms may support ownership transfer or delegated configurability, but often encode this authority indirectly through platform-specific access-control mechanisms. This is sufficient to determine who may reach configuration state, but not whether an authorized update preserves previously established policy boundaries. Bounded reconfiguration therefore requires configuration authority to be represented explicitly and validated against prior protection state. In OB-WC, explicit ownership makes this authority visible and enforceable in hardware, while rule containment provides the bounds within which it may evolve. Encoding configuration authority only through additional protection rules does not provide the same semantics and makes the resulting authority structure harder to reason about.

\mypara{Beyond System-Level Checkers.}
The proposed ownership and reconfiguration semantics are not tied to checker rules and can apply to any configurable resource whose authority may be delegated at runtime. Integrating them into every peripheral may be impractical for legacy or compatibility reasons, making external checkers a convenient enforcement point without modifying protected resources. The same semantics could also be integrated into resource-local access control, as in TrustZone-aware peripherals. Ultimately, the choice between centralized and distributed enforcement depends on system topology and deployment constraints.

\mypara{Hierarchical Trust, not Confidential Computing.}
Our model targets hierarchical trust, where preceding owners may remain part of the TCB. It prevents receivers from escaping established boundaries, but does not require delegators to relinquish access to the delegated resource. This matches the embedded and mixed-criticality MCU systems targeted in this work, where confidential-computing-style protection from more privileged software is not yet a common deployment model. Supporting that stronger threat model would require non-retentive ownership transfer that revokes prior access in hardware, potentially combined with attestation. These guarantees are outside our current scope.

\mypara{Validation Scope Depends on Deployment.}
\textcolor{black}{Our prototype validates each update against the local rule set of a single checker. When checkers protect distinct regions of the physical memory map, each checker can independently validate updates within its region without requiring coherent protection state across checkers. Software must still coordinate policy transitions involving resources protected by multiple checkers, including the ordering of their updates. Alternatively, rules could be maintained centrally and enforced through distributed checkers, enabling system-wide validation while requiring coherent policy propagation across enforcement points.}
%Our prototype validates each update against the local rule set of a single checker. Policies spanning multiple checkers must therefore be coordinated by software. This is an implementation choice rather than a model limitation: rules could instead be maintained centrally and enforced through distributed checkers, enabling system-wide validation while requiring coherent policy propagation across enforcement points.

\mypara{Policy Transitions are not Transactional.}
OB-WC provides validate-before-commit semantics for each configuration request, but does not make multi-rule policy transitions atomic. Software must order updates so that every intermediate committed state preserves the intended security properties. Transactional updates across multiple rules would require additional shadow state or commit-group semantics but they are orthogonal to bounded configuration authority.

\begin{table}[t!]
    \caption{Comparison with related research on system-level protection
    and authority delegation.
    R1: system-level enforcement; R2: transferable authority;
    R3: recursive delegation; R4: bounded delegation.}
    \label{tab:related}
    \centering
    \small
    \renewcommand{\arraystretch}{1.15}
    \setlength{\tabcolsep}{3pt}

    \begin{tabular}{
        >{\centering\arraybackslash}p{2.5cm}
        >{\centering\arraybackslash}p{.55cm}
        >{\centering\arraybackslash}p{.55cm}
        >{\centering\arraybackslash}p{.55cm}
        >{\centering\arraybackslash}p{.55cm}
        >{\centering\arraybackslash}p{2.5cm}}
        \toprule
        \multirow{2}{*}{\textbf{Mechanism}}
        & \multirow{2}{*}{\textbf{R1}}
        & \multirow{2}{*}{\textbf{R2}}
        & \multirow{2}{*}{\textbf{R3}}
        & \multirow{2}{*}{\textbf{R4}}
        & \textbf{Authority Representation} \\
        \midrule

        \multirow{2}{*}{Fiorin et al.~\cite{Fiorin2008}}
        & \multirow{2}{*}{\cmark}
        & \multirow{2}{*}{\xmark}
        & \multirow{2}{*}{\xmark}
        & \multirow{2}{*}{\xmark}
        & Central NSM ownership \\

        \midrule
        \multirow{2}{*}{NoC-MPU~\cite{Porquet2011}}
        & \multirow{2}{*}{\cmark}
        & \multirow{2}{*}{\xmark}
        & \multirow{2}{*}{\xmark}
        & \multirow{2}{*}{\xmark}
        & Trusted software agents \\

        \midrule
        \multirow{2}{*}{Tan et al.~\cite{Tan2017}}
        & \multirow{2}{*}{\cmark}
        & \multirow{2}{*}{\xmark}
        & \multirow{2}{*}{\xmark}
        & \multirow{2}{*}{\halfmark}
        & Static Owner permissions \\

        \midrule
        \begin{tabular}[c]{@{}c@{}}
        CHERI~\cite{Watson2023} \\
        CHERI-TrEE~\cite{Strydonck2023} \\
        CAPSTONE~\cite{Yu2023}
        \end{tabular}
        & \halfmark
        & \cmark
        & \cmark
        & \cmark
        & \begin{tabular}[c]{@{}c@{}}
        Capabilities / \\ 
        derivation
        \end{tabular} \\

        \midrule
        CapChecker~\cite{cheri_capchecker}
        & \cmark
        & \halfmark
        & \halfmark
        & \cmark
        & CHERI capabilities \\

        \midrule
        \multirow{2}{*}{sIOPMP~\cite{Feng2024}}
        & \multirow{2}{*}{\cmark}
        & \multirow{2}{*}{\halfmark}
        & \multirow{2}{*}{\halfmark}
        & \multirow{2}{*}{\cmark}
        & Capabilities + monitor \\

        \midrule
        \multirow{2}{*}{\small \textit{\textbf{OB-WC}}}
        & \multirow{2}{*}{\cmark}
        & \multirow{2}{*}{\cmark}
        & \multirow{2}{*}{\cmark}
        & \multirow{2}{*}{\cmark}
        & \textbf{Ownership + protection state} \\

        \bottomrule
    \end{tabular}

    \cmark~Full support,
    \halfmark~Partial, indirect, or statically constrained support,
    \xmark~No support. \\
    \vspace{1ex}
    \textit{For Tan et al., bounded delegation applies to dynamically granted resource-access permissions, while Owner authority remains statically assigned. For capability-based systems, R2–R4 refer to capability/resource authority rather than authority over rule-based protection configuration.}
\end{table}

\section{Related Work}

Table~\ref{tab:related} compares OB-WC with prior research on system-level protection and capability-based authority in how configuration authority is represented, transferred, and bounded.

\mypara{System-Level Dynamic Protection.}
Prior work on system-level protection has explored dynamic access control using distributed enforcement points. Fiorin et al.~\cite{Fiorin2008} place programmable protection units at NoC interfaces, with a centralized security manager retaining ownership and mediating reconfiguration. NoC-MPU~\cite{Porquet2011} similarly enforces compartment-level permissions across heterogeneous initiators, while configuration remains under trusted software control. These approaches support dynamic resource policies but do not delegate configuration authority to the domains whose resources they protect. Tan et al.~\cite{Tan2017} come closest to our containment model. Their Isolation Units allow a task to dynamically share only subregions covered by a statically assigned Owner permission. However, Owner authority itself is fixed at design time and cannot be transferred or recursively delegated: receivers obtain resource-access authority, not configuration authority. OB-WC instead makes rule ownership transferable and allows receivers to further delegate contained rules, while enclosing rules preserve previous bounds.

\mypara{Capability-Based Authority Delegation.}
Capability-based architectures provide a complementary approach in which authority and its bounds are explicitly represented by capability objects and derivation relationships. seL4~\cite{Klein2009} uses
capabilities with attenuated rights and explicit derivation relationships, while CHERI~\cite{Watson2023} provides architecturally enforced bounds and permissions. CHERI-TrEE~\cite{Strydonck2023} and CAPSTONE~\cite{Yu2023} further support dynamic, nested, and delegated authority. \textcolor{black}{Markettos et al.~\cite{Markettos2020} propose capability-based DMA protection, including capability search units whose configuration may be delegated to applications within their existing capability bounds. 
CapChecker~\cite{cheri_capchecker} subsequently constrains} capability-unaware accelerators through an external enforcement interface. sIOPMP~\cite{Feng2024} similarly uses capabilities managed by a trusted monitor to authorize system-level I/O mappings. OB-WC differs in how authority is represented and enforced. It uses no separate capability or derivation structure. Instead, ownership identifies who may reconfigure a rule, while containment in the active protection state bounds how that authority may be refined or delegated. This targets lightweight rule-based protection mechanisms common in MCU-class SoCs and remains compatible with legacy software.

\section{Conclusion}
We presented bounded configuration authority, a hardware-supported model for securely evolving protection policies at runtime in next-generation MCUs. Our RISC-V World Checker implementation demonstrates that these semantics can be enforced off the normal access datapath with modest hardware overhead, while significantly reducing policy-transition latency compared with trusted software mediation. \textcolor{black}{Together, these results demonstrate that lightweight protection hardware can give software the autonomy to evolve protection policies while preserving the boundaries that make this autonomy safe. The proposed mechanisms can inform the architectural evolution and ongoing standardization of RISC-V Worlds.}

% \section*{Acknowledgements}
% This document is derived from previous conferences, in particular ISCA 2019,
% MICRO 2019, ISCA 2020, MICRO 2020, ISCA 2022, HPCA 2022, and ISCA 2024. 

%%%%%%% -- PAPER CONTENT ENDS -- %%%%%%%%

%%%%%%%%% -- BIB STYLE AND FILE -- %%%%%%%%
\bibliographystyle{IEEEtranS}
\bibliography{refs}
%%%%%%%%%%%%%%%%%%%%%%%%%%%%%%%%%%%%

\end{document}